\documentstyle[epsfig]{aipproc}

\begin{document}

\title{TGRS OBSERVATIONS OF POSITRON ANNIHILATION IN CLASSICAL NOVAE}

\author{M. J. Harris$^{1}$, D. M. Palmer$^{1}$, J. E. Naya$^{1}$, B. J. 
Teegarden, T. L. Cline, N. Gehrels, R. Ramaty, H. Seifert$^{1}$}
\address{Code 661, NASA/Goddard Spaceflight Center, Greenbelt, MD 20771\\
$^{1}$ Universities Space Research Association}

\maketitle

\begin{abstract}

The TGRS experiment on board the {\em Wind} spacecraft has many advantages
as a sky monitor  ---  broad field of view ($\sim 2 \pi$
centered on the south ecliptic pole), long life (1994--present), and
stable low background and continuous coverage due to {\em Wind\/}'s
high altitude high eccentricity orbit.  The Ge detector has sufficient
energy resolution (3--4 keV at 511 keV) to resolve a cosmic positron
annihilation line from the strong background annihilation line from
$\beta$-decays induced by cosmic ray impacts on the instrument, if the
cosmic line is Doppler-shifted by this amount.  Such lines (blueshifted)
are predicted from nucleosynthesis in classical novae.  We have searched
the entire TGRS database for 1995--1997 for this line, with negative
results.  In principle such a search could yield an unbiased upper
limit on the highly-uncertain
Galactic nova rate.  We carefully examined the times around the known
nova events during this period, also with negative results.  The
upper limit on the nova line flux in a 6-hr interval is
typically $<3.8 \times 10^{-3}$
photon cm$^{-2}$ s$^{-1}$ ($4.6 \sigma$).  We performed the same
analysis for times around the outburst of Nova Vel 1999, obtaining a
worse limit due to recent degradation of the detector response
caused by cosmic ray induced damage.

\end{abstract}

\section{INTRODUCTION}

Theoretical models of classical novae imply that large quantities of
$\beta$-unstable proton-rich nuclei are formed during a thermonuclear
runaway, with half-lives of the order minutes to hours.  The resulting
positrons undergo annihilation in the expanding nova envelope, giving
rise to a pulse of annihilation $\gamma$-rays which
is blueshifted and broadened by the nova velocity and lasts for
up to $\sim 6$ hr (2).

All space-borne instruments are subject to a strong background 
emission line at 511 keV, arising from annihilation of 
cosmic ray induced $\beta$-decay positrons in the instrument,
which hinders detection of the same line from cosmic sources.  However,
a spectrometer with sufficient energy resolution could distinguish
between cosmic and background lines because the former is blueshifted,
typically by 2--5 keV.  The Ge spectrometer TGRS (FWHM resolution
3 keV at the beginning of its mission) fulfills this requirement.
Its other advantages as a detector and monitor of the nova 511 keV
line are its broad aperture (nearly $2 \pi$ sterad) and its
high altitude high inclination orbit, in which  
background count rates are rather low, due to the
lack of interaction with Earth's trapped
radiation belts, and to the virtual absence of Earth albedo background
radiation.  The background count rate and spectrum has also been
extremely stable throughout the mission.

\begin{figure}
\centerline{\epsfig{file=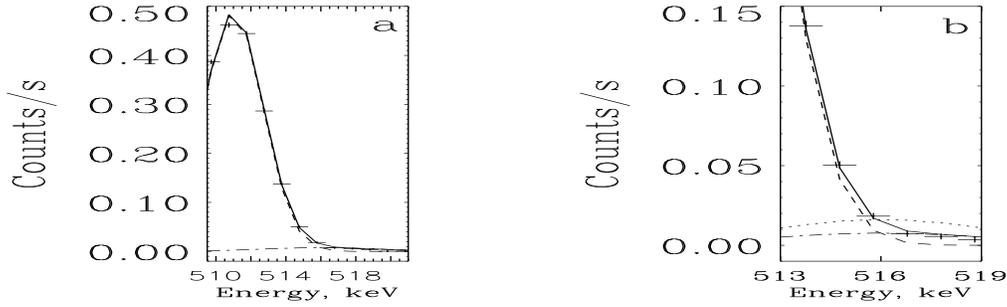,height=2.0in,width=6.0in}}
\vspace{10pt}
\caption{Characteristic TGRS background count spectrum at energies around
511 keV, obtained during the
interval 3 June 12h--18h UT with a power law continuum subtracted.  The
components of the fit are a line with the 
width and position of the nova 6-hr line (2, dot-dashed line) and a 
Gaussian line fitting the blue wing of the
511 keV background line (dashed line).  The total model spectrum
is the full line.  (b)  Expansion of Fig. 2a showing the significance of the
fitted nova 6-hr line (dot-dashed line of amplitude $5.2 \pm 0.1 \times
10^{-3}$ photon cm$^{-2}$ s$^{-1}$; other symbols as in Fig. 2a).
Also shown (dotted line) is the theoretically expected level of the 
nova 6-hr line for a nova at 1 kpc (ref. 6 model HH5).}
\label{Fig. 1}
\end{figure}

A disadvantage of Ge detectors for long-term missions is progressive
degradation of performance due to cosmic-ray impact damage.  
For this reason most of the
data used here were obtained during 1995--1997, before this effect
became serious.  In an attempt
to avoid the effects of degradation, such as low-energy "tailing",
we fitted only energies on the high-energy (blue) side of the
background 511 keV line.  One such fitted background spectrum 
is shown in Fig. 1.  The
dashed line is the strong background 511 keV line.  The dot-dashed
line is the fit to a theoretical nova line emitted in a 6-hr interval
after the event (3).  Although
the amplitude of this line is significant, we will see shortly that
it is characteristic of all our background spectra --- in other words,
there is a systematic constant positive offset in this measurement.
For scale, the dotted line shows how the same line would appear
from a model nova at a distance 1 kpc.  This line would be
detected at a level $\sim 8 \sigma$.

\section{SURVEY OF THE SOUTHERN SKY, 1995--1997}

The global rate of classical novae in the Galaxy is highly uncertain,
since the two methods used to derive it both suffer from
serious systematic problems.  One involves correcting the 
rate of actual discoveries ($\sim 3$ yr$^{-1}$) for highly
uneven sky coverage and interstellar extinction for poorly-known
nova distances.  The other involves correcting the nova
rates measured in external galaxies (according to some parameter such as
$M_{B}$), which runs into the problem of differences in the stellar
populations between galaxies.

The method used here is potentially superior, being free from
these biases.  The 511 keV $\gamma$-rays are not subject to interstellar
extinction, and TGRS's sky coverage was almost uniform during the period
1995--1997.  Between 1995 January and 1997 October TGRS
accumulated about $7.7 \times 10^{7}$ s of background spectra (88\%
overall).  Nor are much data lost due to changing sensitivity across the
aperture, which amounts to a factor $\sim$25\% between the zenith and
the ecliptic plane.

This entire period was divided into 6 hr intervals during which background
spectra were accumulated, which were fitted to models of the type
shown in Fig. 1.  The parameters of the nova line in these fits were
fixed at values predicted by theory at epochs $\simeq 6$ hr and
$\simeq 12$ hr after the explosion (widths 8 keV FWHM, blueshifts 5 and 2
keV, from ref. 2 and Hernanz 1997, private communication).  The
fits to the count spectra were normalized by the effective area for the
zenith angle of an average nova, which is $\simeq 60^{\circ}$.  The 
fitted amplitudes for the 6-hr and 12-hr lines were combined in order to
maximize the signal.

\begin{figure}
\centerline{\epsfig{file=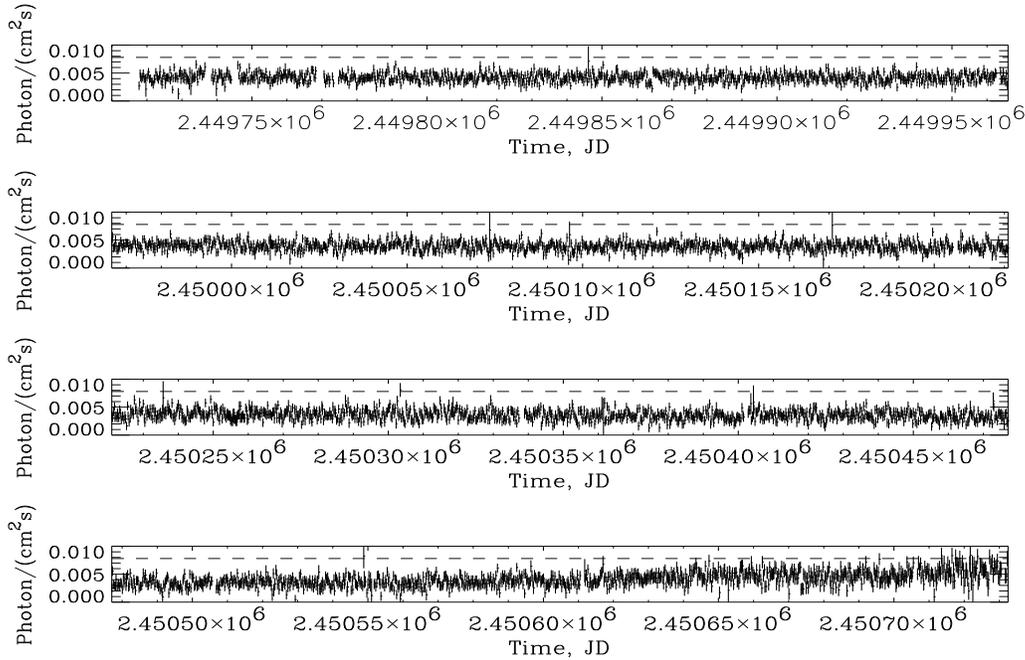,height=3.5in,width=5.6in}}
\vspace{10pt}
\caption{Measured fluxes in the nova 6-hr line during the entire interval
1995 January -- 1997 October.  Dotted line --- mean $4.6 \sigma$ limit,
above which candidate line detections would lie.}
\label{Fig. 2}
\end{figure}

The fitted nova line amplitudes in the 4005 resulting background spectra are 
shown in Fig. 2.  Since the positive systematic amplitude is constant to a 
very good approximation, we are justified in subtracting it from
all measurements.  We then see that there is no single 6-hr interval which, 
after this subtraction, yields
a convincing nova signal.  We chose a level $4.6 \sigma$ as the threshold
for detection of a candidate signal (Fig. 2, dashed line), which is appropriate
for the expected number of chance detections in 4005 events.  From Poisson
statistics, our measured nova rate is therefore $< 1$ at the
63\% level.  The sensitivity of this survey at the $4.6 \sigma$ level is about
$3.8 \times 10^{-3}$ photon cm$^{-2}$ s$^{-1}$ (4).

If a given nova model predicts a 511 keV line flux $\phi_{pred}$ at 1 kpc
then this survey could have detected it out to a distance $r_{det} 
= \sqrt{\phi_{pred}(M)/3.8 \times 10^{-3}}$ kpc.  Unfortunately the most
recent self-consistent hydrodynamic models (5) predict much smaller
fluxes ($< 2 \times 10^{-3}$ photon cm$^{-2}$ s$^{-1}$ at 1 kpc) than 
earlier versions (2), although parametrized models (6) can predict values 
values as high as $7 \times 10^{-3}$ photon cm$^{-2}$ s$^{-1}$.  Thus
in the most optimistic case $r_{det} = 1.4$ kpc, for an ONeMg nova of model
type HH5 (6).  Using the Bahcall-Soneira Galactic mass model (1)
we can extrapolate our limit $< 1$ event per $7.7 \times 10^{7}$ s 
within 1.4 kpc (which includes $\sim$0.8\% of the Galaxy) to
a Galactic rate $<54$ events yr$^{-1}$ of this type.

\section{NOVA VELORUM 1999}

Five novae occurred within TGRS's aperture during the period 1995--1997,
before the instrument performance began to deteriorate.  Searches for
511 keV line emission around the times of these outbursts were unsuccessful
(3), the sensitivities achieved being similar to the values given above
for the overall survey.  Unfortunately a very bright nova V382 Vel 
($\equiv$ N Vel 1999) was discovered in 1999 May, by which time the
deterioration of the detector performance had advanced. 

Three main effects contribute to this deterioration (7): a 
distortion of the response to a narrow line
(starting with low-energy tailing, culminating in the formation of
a separate low-energy peak); a poorly-quantified loss of efficiency; and
an accelerated rate of change of the instrument gain.  In order to apply
our search method to Nova Vel we attempted approximate solutions to
these problems.  In the count
spectrum fitting, the Gaussian line models (Fig. 1) were replaced by
more complex shapes reflecting the current detector response (evaluated
from the response to the 511 keV and other background lines).  
Fortunately, during the rather narrow time interval (15--23 May) which we
fitted the response did not change significantly, nor did the gain.  
We estimated the detector effective area by assuming constancy of the true TGRS
background line strengths, which we had found to be a very good approximation
during the period of stability 1995--1997;
a measurement of the 511 keV line intensity, compared with the average
from 1995--1997, then enabled an efficiency correction to be made.  

The preliminary results are shown in Fig. 3.  Note that, compared to
Fig. 2, both the error bars and the systematic positive offset are much
worse.  The rise to visual maximum probably started at least 1 d before
discovery (arrow), and the explosion may have occurred up to 3 d before
that.  Our fitted background nova line amplitude
should show a single 6-hr peak at that epoch, over and above the average
background level (solid line).
No such event was observed, so that we can only place upper limits on
the line flux and a lower limit on the distance to V382 Vel.  The average
$3 \sigma$ upper limit above the solid line is $<1.3 \times
10^{-2}$ photon cm$^{-2}$ s$^{-1}$; however, the constant positive offset
level is much less well-defined than it was earlier in the mission (Fig.
2), so that a systematic error estimate for this must be introduced.
We also estimated the systematic error due to the uncertainty in the
nova line position arising from the broad and irregular shape of the
background 511 keV line relative to which it is measured.
We thus obtained an upper limit on the line flux of order $<1.6 \times
10^{-2}$ photon cm$^{-2}$ s$^{-1}$.  V382 Vel is believed to have
occurred on an ONeMg white dwarf (8), in which 
case we can combine our flux limit with the best-case theoretical flux 
at 1 kpc of $7 \times
10^{-3}$ photon cm$^{-2}$ s$^{-1}$ (6) to obtain a lower limit of 
0.66 kpc on the distance.

\begin{figure}
\centerline{\epsfig{file=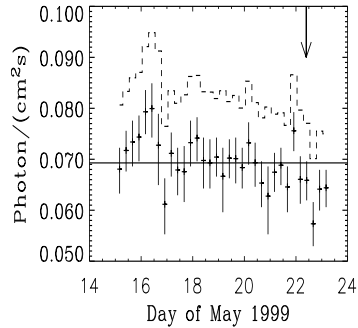,height=2.0in,width=2.0in}}
\vspace{10pt}
\caption{Measured fluxes at 6-hr intervals prior to the discovery of
V382 Vel (arrow).  Full line --- assumed constant positive offset.
Dashed line --- $3 \sigma$ upper limits on the line flux.}
\label{Fig. 3}
\end{figure}

\end{document}